\documentclass[aps,prl,preprint,groupedaddress]{revtex4-1}

\usepackage{graphicx}
\usepackage{color}
\usepackage{dcolumn}
\usepackage{amsmath}
\usepackage{amssymb}
\usepackage{amsfonts}
\usepackage{bm}
\usepackage{epstopdf}
\usepackage{picture}
\usepackage{enumitem}
\usepackage{afterpage}
\usepackage{placeins}
\setlist[enumerate]{label*=\arabic*.}

\usepackage{xr}
\newcommand{\Fvec}{\mathbf{F}}
\newcommand{\fvec}{\mathbf{f}}

\newcommand{\rvec}{\mathbf{r}}
\newcommand{\Evec}{\mathbf{E}}
\newcommand{\Hvec}{\mathbf{H}}

\newcommand{\Svec}{\mathbf{S}}

\newcommand{\pvec}{\mathbf{p}}
\newcommand{\kg}{k_r}
\newcommand{\ks}{k_{\phi}}

\newcommand{\Kg}{K_r}
\newcommand{\Ks}{K_{\phi}}
\newcommand{\Pthr}{P_0}

\begin{document}

\title{Orbital Motion From Optical Spin: The Extraordinary Momentum Of Circularly Polarized Light Beams}

\author{V. Svak, O. Brzobohat\'{y}, M.\v{S}iler, P. J\'{a}kl, J. Ka\v{n}ka, P. Zem\'{a}nek, and S. H. Simpson}
\email{simpson@isibrno.cz}
\affiliation{The Czech Academy of Sciences, Institute of Scientific Instruments, Kr\'{a}lovopolsk\'{a} 147, 612 64 Brno, Czech Republic}


\begin{abstract}
We provide a vivid demonstration of the mechanical effect of transverse spin momentum in an optical beam in free space. This component of the Poynting momentum was previously thought to be virtual, and unmeasurable. Here, its effect is revealed in the inertial motion of a probe particle in a circularly polarized Gaussian trap, in vacuum. Transverse spin forces combine with thermal fluctuations to induce a striking range of non-equilibrium phenomena. With increasing beam power we observe (i) growing departures from energy equipartition, (ii) the formation of coherent, thermally excited orbits and, ultimately, (iii) the ejection of the particle from the trap.
Our results complement and corroborate recent measurements of spin momentum in evanescent waves, and extend them to a new geometry, in free space. In doing so, we exhibit fundamental, generic features of the mechanical interaction of circularly polarized light with matter. The work also shows how observations of the under-damped motion of probe particles can provide detailed information about the nature and morphology of momentum flows in arbitrarily structured light fields as well as providing a test bed for elementary non-equilibrium statistical mechanics. 

\end{abstract}
\maketitle
Although its existence has been recognized since the time of Kepler, the momentum of light remains a subject of intense debate. Uncertainty over its value when propagating through continuous media continues to excite interest \cite{Pfeifer2009Colloquium}. Even in free space the topic is not without controversy. For example, the time averaged Poynting momentum in free space can be separated into two parts \cite{Berry2009Optical,Bekshaev2011Internal,Neugebauer2015Measuring} whose differing properties have yet to be fully understood:

\begin{equation}
\pvec = \frac{1}{2c^2} \Re \left( \Evec^{\ast} \times \Hvec \right) = \frac{1}{2c^2}\Im \left[\Evec^{\ast} \cdot (\nabla) \Evec \right] + \frac{1}{4c^2} \nabla \times \Im \left[\Evec^{\ast} \times \Evec \right] \equiv 
\pvec^{O} + \pvec^{S},
\end{equation}
 (in SI units, with $c$ the speed of light in vacuum). The first of these contributions, $\pvec^O$, is independent of polarization, and referred to as the orbital component. The second term, $\pvec^S$, is related to inhomogeneous circular polarization, and referred to as the spin component.

\begin{figure}[h!]
	\includegraphics[width=0.9\textwidth]{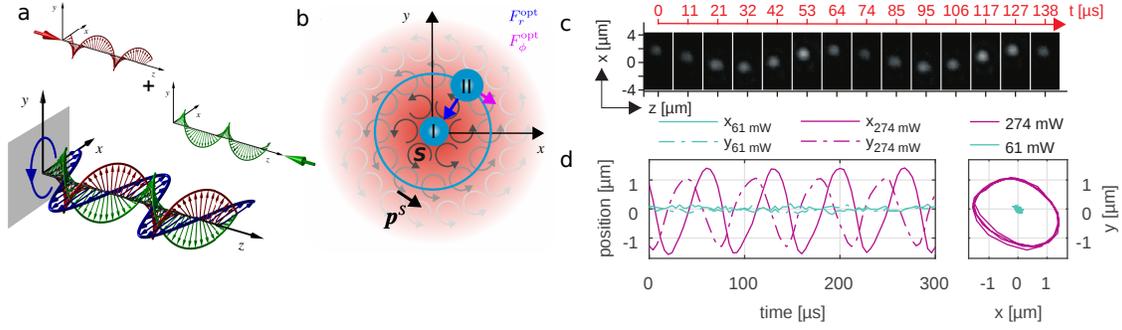}
	\caption{\label{fig:beam+exp} Physical principle. (a) Schematic showing the electric fields in the counter-propagating circularly polarized beams of opposing handedness (green and red) and the rotation of the combined electric field (blue). (b) Local optical spin ($\Svec=\Im(\Evec^{\ast} \times \Evec)$) field in the transverse xy plane of the beam. Radial gradient optical force $F^{opt}_r$ and azimuthal non-conservative spin force, $F^{opt}_{\phi}$ acting upon a particle positioned off-axis. Regime I corresponds to the particle position in the vicinity of the beam axis for lower laser power while regime II corresponds to the above threshold condition, in which the particle orbits the beam axis. (c) Snap shots of orbiting particle (regime II) taken by the CCD camera oriented perpendicularly to the beam propagation. (d) Trajectories of the particle for both regimes I and II acquired by QPD. Turquoise curves denote trajectories for lower trapping power 55 mW and particle motion near the beam axis; pink trajectories show an orbiting particle for a power of 180 mW, above threshold.}
\end{figure}

Interest in these distinct contributions is rooted in field theory, as has been extensively discussed elsewhere \cite{Bliokh2014Extraordinary,Bliokh2013Dual,Bliokh2011Spin-to-orbital}. In summary, the field theoretic basis of the Poynting momentum rests on Belifante's symmetrization of the canonical stress energy tensor, derived by applying Noether's theorem to the electromagnetic Lagrangian \cite{Jackson1999Classical,Soper2008Classical}. This modification ensures conservation of angular momentum, removes the gauge dependence of physical quantities and resolves various other inconsistencies \cite{Jackson1999Classical,Soper2008Classical}. In this context, $\pvec^O$ can be identified as the canonical momentum derived from the Noether theorem, evaluated in the Coulomb gauge. This term gives the optical scattering force on a point particle (Supplementary Information). The spin part of the Poynting vector, $\pvec^S$, appears as a consequence of symmetrization and was thought to be virtual, since it does not contribute to the overall momentum of the field ($\int_{\infty} \pvec^S dV = 0$). The physical origin of $\pvec^S$ can be conceptually understood in terms of spin momentum loops \cite{Bliokh2014Extraordinary}. In infinite, homogeneous fields, contributions from neighbouring loops cancel. Spatial inhomogeneities, including intensity gradients, break this balance, generating a non-zero boundary spin current, $\pvec^S \neq 0$ \cite{Bliokh2014Extraordinary}. Figure (\ref{fig:beam+exp}b) illustrates this mechanism for circularly polarized beams, as used in this work. \\

The question of how and, indeed, if $\pvec^S$ couples to matter presents itself. Addressing this issue is experimentally challenging. Spin momentum ($\pvec^S$) requires field inhomogeneities which, by necessity, generate other forms of optical force, including gradient forces. To measure its effect, delicate spin forces must be reliably distinguished from other, stronger forces. 
This was elegantly achieved in a recent experiment, in which the deflection of a nano-cantilever, immersed in a circularly polarized evanescent field was measured \cite{Antognozzi2016Direct}. In this case the spin dependence of the force is isolated, from the prevailing effects of orbital momentum, by varying the degree of circular polarization, applying a symmetry argument and by observing that, in an evanescent wave, spin and orbital components of momentum are orthogonal \cite{Bliokh2014Extraordinary,Bliokh2014Extraordinary}. Other experiments require minute analysis of the motion of particles in liquid, subject to appropriate light fields \cite{Angelsky2012Orbital,Ruffner2012Optical}. Related or analogous effects rely on substrate interactions or optical chirality or birefringence \cite{Sukhov2015Dynamic,Rodriguez-Fortuno2015Lateral,Wang2014Lateral,Magallanes2018Macroscopic}
and, whilst fascinating on their own right, do not depend explicitly on incident spin momentum.\\

In our experiments, spin forces manifest themselves directly, through the inertial motion of a microsphere in a vacuum optical trap, in free space. Such traps are used in the pursuit of ground state cooling of macroscopic objects \cite{Li2011Millikelvin,Chan2011Laser,Millen2015Cavity}, for studying aspects of nano-scale dynamics \cite{Gieseler2014Nonlinear,Arita2013Laser} and statistical mechanics \cite{Rondin2017Direct,Millen2014Nanoscale,Gieseler2014Dynamic}. 
By comparison with the relatively subtle effects previously observed, our results are striking and unequivocal. Ultimately, we observe the violent ejection of a probe particle from an optical trap, as a direct consequence of transverse optical spin momentum. The experiment itself is elegant in its simplicity. We track the motion of a small, dielectric probe particle in a counter-propagating Gaussian beam trap, in vacuum (see Fig. (\ref{fig:beam+exp}a,b)). The beams are weakly focused with a numerical aperture of $NA \approx 0.18$ (equivalent to a Rayleigh range of $kz_R \approx 150$, for wave-vector $k$). This symmetric geometry nullifies the axial scattering forces associated with each beam \cite{Divitt2015Cancellation} and emphasises transverse spin components, when they are present. The polarization state of the beams can be independently adjusted using quarter wave plates on either side of the trap (see Methods). When the beams have parallel, linear polarization (LP), the only surviving forces in the trap are optical gradient forces. These forces attract the particle towards the beam axis, and provide a periodically modulated force, in the axial direction, that is associated with the interference fringes of the standing wave (see Supplementary Information). Since gradient forces are conservative thermal equilibrium is maintained within the trap and the thermal motion of the particle complies with the equipartition of energy, so that elastic energy of the trap is equal to the thermal energy. In particular, the optical restoring force, $\Fvec_i$, is linearly proportional to displacement, $x_i$, to first order e.g. $F_i=-K_i x_i$, where $K_i$ is the trap stiffness in direction $i$. $K_i$ is directly proportional to the optical power, $K_i=Pk_i$. Equating thermal and elastic energies gives $\frac{1}{2}k_BT=\frac{1}{2} K_i \langle x_i^2 \rangle$, so that the greater the power in the trap, the more tightly the particle is constrained i.e. $\langle x_i^2 \rangle \propto 1/P$.\\

When the quarter-wave plates are adjusted so that the beams are circularly polarized with opposite handedness (CP) the electric field vectors of the beams rotate together, in the same sense, Fig. (\ref{fig:beam+exp}a). This has two effects. First, the weak focusing by the lens produces small, transverse components of $\pvec^O$ through spin-orbit coupling \cite{Nieminen2008Angular}. Second, the inhomogeneous circular polarization generates transverse components of $\pvec^S$, which also circulate about the beam axis, Fig. (\ref{fig:beam+exp}b). The spin term is strongly dominant, with $\pvec^S/\pvec^O \gtrsim (2k z_R) \approx 300$ (Supplementary Information). These transverse momenta give rise to azimuthal forces, primarily associated with $\pvec^S$ (Supplementary Information). In combination with the optical gradient forces, the circularly polarized trap has a cylindrically symmetric force field, $\Fvec^{\mathrm{opt}} = F_{\phi}^{\mathrm{opt}} {\bf e}_\phi + F_r^{\mathrm{opt}} {\bf e}_r$, with all forces vanishing on the beam axis (see Supplementary Information). For small displacements from the axis the gradient force, $F^{opt}_r$ is restoring, attracting the particle towards the center of the beam. The azimuthal component, $F_{\phi}$, makes the trap non-conservative (i.e. $\nabla \times \Fvec^{\mathrm{opt}} \neq 0$), forcing it out of equilibrium and allowing the particle to accumulate angular momentum. Notably, the ratio of the radial gradient force to the azimuthal spin force is approximately constant over the width of the beam i.e. $F^{opt}_{\phi}/F^{opt}_r \approx$ const, because they have a similar dependence on intensity gradients (see Supplementary Information). A detailed theoretical analysis of the momentum flows and optical forces acting on small and Mie particle is provided in Supplementary Information.

The motion of the probe particle is determined by the delicate interplay between thermal fluctuations, spin and gradient optical forces and is governed by a Langevin equation, in Cartesian coordinates:

\begin{equation} \label{eq:Lang}
\Fvec^{\mathrm{opt}}(\rvec) + \Fvec^{\mathrm L}(t) - \xi \dot \rvec = m \ddot \rvec.
\end{equation}
Where $m$ is the mass of the particle and $\xi$ the pressure dependent Stokes' drag. $\Fvec^{\mathrm{opt}}\equiv P\fvec^{opt}$ is the external optical force at power $P$ and $\Fvec^{\mathrm L}$ is the fluctuating Langevin force, uncorrelated in time with zero mean and variance given by the fluctuation-dissipation theorem e.g. $\langle F^L_i(t) \rangle=0$, $\langle F_i^L(t) \otimes F_j^L(t') \rangle=2k_BT\xi \delta_{ij} \delta(t-t')$ for Cartesian components $i,j=x,y,z$. For circularly polarized beams we observe two distinct regimes (see depiction in Fig. (\ref{fig:beam+exp}c,d)), dependent on optical power. These can be contrasted with the conspicuously different behaviour observed in linearly polarized traps, which is qualitatively similar for all trap powers. At low optical powers (\textbf{regime I}) the particle undergoes non-equilibrium Brownian motion, deviating substantially from the equipartition of energy. At higher powers (\textbf{regime II}) we observe formation of stable orbits, which are thermally excited and driven by optical spin.\\

\section{Regime I: Driven Brownian Motion}
In the {\bf low power regime I} (turquoise curves, Fig. (\ref{fig:beam+exp}d)), the particle remains within the linear range of the trap and, in Cartesian coordinates, the optical force can be approximated as, 
\begin{equation}\label{eq:stiff}
\begin{bmatrix} F_x^{\mathrm{opt}} \\ F_y^{\mathrm{opt}} \end{bmatrix} =
- \begin{bmatrix} \Kg & \Ks \\ -\Ks & \Kg \end{bmatrix} \begin{bmatrix} x \\ y \end{bmatrix}.
\end{equation}
$\Kg$ is the radial stiffness associated with the gradient force. It is isotropic and gives a force in the radial direction, towards the origin. $\Ks$ is the stiffness of the azimuthal force. Both stiffness coefficients are proportional to power e.g. $\Kg=P\kg$ and $\Ks=P\ks$.\\
The resulting linear system has a set of characteristic frequencies $\{ \omega_c \}$, that relate to qualitative features of the stochastic dynamics. In the case of weak damping, $\xi^2 \ll 4 m \Kg$ (Supplementary Information )
:
\begin{equation}
\omega_{c} \approx 
\pm \omega_0+\frac{i}{2m} 
\left(\xi \pm  \frac{\Ks}{\omega_0} \right).
\end{equation}
The real parts of $\omega_c$ describe typical oscillation frequencies and, for all polarizations, they are approximately equivalent to the resonant trap frequency, $\omega_0 = \sqrt{P\kg / m}$. The imaginary parts of $\omega_c$, $\Im(\omega_c)$, represent damping or loss coefficients, associated with exponential decay. For linearly polarized beams the azimuthal force is zero, $\ks=0$ so that $\Im(\omega_c) = \xi/m$ is positive and independent of power. In contrast, when $k_{\phi} \ne 0$, the imaginary parts of two of the characteristic frequencies approach zero as the optical power approaches a critical value, $P_0$: 
\begin{equation}
\frac{P_0}{\xi^2}=\frac{k_r}{m k_{\phi}^2}.
\label{eq:bal1}
\end{equation} 
At $P=P_0$, the corresponding threshold frequency $\Omega_0\equiv\omega_0(P_0)$ satisfies,  
\begin{equation}
\frac{\Omega_0}{\xi}=\frac{k_r}{m k_{\phi}}.
\label{eq:bal2}
\end{equation}
As demonstrated below, these conditions (Eqs. (\ref{eq:bal1}, \ref{eq:bal2})) describe the boundary between regimes I and II, and arise when average centripetal forces acquired by the Brownian particle are just sufficient to balance attractive, gradient forces.

In regime I, where the power is below threshold ($P<P_0$), the stochastic motion of the probe particle can be characterised in terms of the power spectral density (PSD) and the time correlations of the particle coordinates. The former quantifies the energy in the particle motion as a function of frequency. The latter 
relates to underlying deterministic motion. For instance, $\langle x(t) y(t+\tau) \rangle$ measures the correlation between the $x$ coordinate at time $t$ with the $y$ coordinate at a later time $(t+\tau)$. For $\tau=0$ we get the instantaneous variances of the coordinates. Both the PSD and the time correlations can be expressed in terms of $\omega_c$ (see Supplementary Information ) and both depend qualitatively on the presence of spin forces (i.e. on whether or not $\ks=0$). Table (\ref{tab:quants}) summarises the results.\\

\begin{table}[h!]
\centering
\begin{tabular}{l | c | c }
 & Linear polarization (LP)
 & Circular polarization (CP)
 \\ 
Quantity 
& $K_\phi=0$ 
& $K_\phi \ne 0$ 
\\ \hline 
$\omega_c$
& $\approx\pm \omega_0+\frac{i \xi}{2m}$ 
& $\approx\pm \omega_0+\frac{i \xi}{2m} \left(1\pm \sqrt{\frac{P}{P_0}}\right)$ 
\\\hline
PSD peak height 
& decreases with power
& increases with $P\rightarrow P_0$ 

\\ 
$\langle X(\omega) X^{\ast}(\omega) \rangle_{max}$ 
& $ =\frac{2k_B T}{\xi \omega_0^2}  \propto \frac{1}{P}$
& $ \approx \frac{ 2k_B T}{m^2} \frac{(\xi^2 \omega_0^2 + \Ks^2)P_0^2}{\xi \omega_0^2 (P-P_0)^2} \propto \frac{1}{(P-P_0)^2}$
\\
\hline 
PSD peak width 
& power independent 
&  decreases towards zero with increasing power
 \\ 
at half maximum 
& $\approx\frac{\xi}{2 m}$ 
& $\approx\frac{\xi}{2m} \left(1\pm \sqrt{\frac{P}{P_0}}\right)$
 \\
\hline 
Time correlations 
& decay power independent
& decay decreases with $P\rightarrow P_0$   
\\
$\langle x(t)x(t+\tau) \rangle$
& $\approx \frac{k_B T}{\Kg} 
\cos(\omega_0 \tau)
e^{-\xi \tau / 2m}$
& $ \approx  \frac{\cos(\omega_0 \tau)}{P-P_0}  e^{\xi (P-P_0) \tau / 4 m \Pthr} $
\\ 
$\langle x(t+\tau) y(t) \rangle $ 
& & $ \approx  \pm \frac{\sin(\omega_0 \tau)}{P-P_0} e^{\xi (P-P_0) \tau / 4 m \Pthr} $ \\
\hline
\end{tabular}
\caption{\label{tab:quants} Characteristic frequencies, $\omega_c$, PSD characteristics and time correlations describing under-damped stochastic motion in circularly and linearly polarized beams}
\end{table}

In both cases, the power spectral density of the system is dominated by a single peak, approximately at the resonant frequency, $\omega_0 = \sqrt{P\kg/m}$ (see Supplementary ). For conservative, linearly polarized beams, the peak height decreases with increasing optical power, $\propto 1/P$ and has a constant width associated with the drag term.

\begin{figure}
\includegraphics[width=0.9\textwidth]{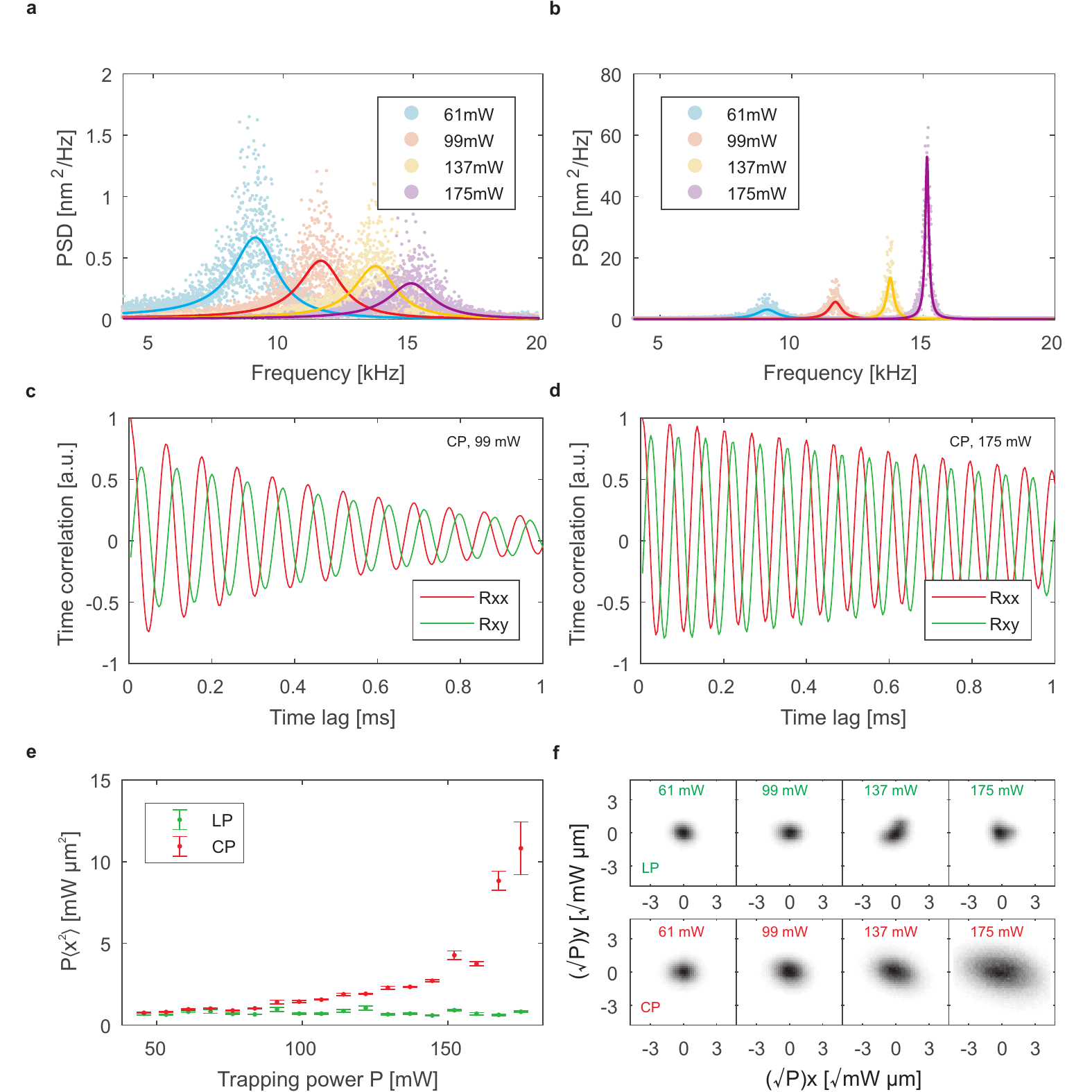}
\caption{\label{fig:subt} Power spectral densities for particle positions for (a) linearly and (b) circularly polarized beams. In the latter case, spin momentum drives a resonance, increasing the peak height and narrowing its width.  (c) and (d) time dependent variance of particle coordinates at two different beam powers. $\langle x(t+\tau)x(t) \rangle$ and $\langle x(t+\tau)y(t) \rangle$ are $\pi/2$ phase shifted, indicating a tendency towards circular motion. Increasing the power increases the mean frequency of rotation and increases the time constant governing the loss of coherence of the oscillation. (e) Graph showing the product of the position variance with power, as a function of power ($P \langle x^2 \rangle$ v $P$), for LP and CP. 
(f) Measured probability distribution of the trapped particle in the transverse plane, scaled by the square root of the beam power for a linearly polarized trap (top row) and a circularly polarized trap (lower row). Variation of this distribution with beam power indicates deviation from thermodynamic equilibrium. 
}
\end{figure}

For very low optical power, the circularly polarized trap behaves similarly. However, as the power increases and approaches the threshold, $P_0$, a qualitative difference emerges and the peak grows in height, $\propto 1/(P_0-P)^2$, while decreasing in width in proportion to $\Im(\omega_c) \approx \frac{\xi}{2m}\frac{(P_0-P)}{2P_0}$. Physically, the white thermal noise excites a resonance in which frictional losses are increasingly compensated by the non-conservative spin forces and the behaviour becomes increasingly coherent and deterministic. In this respect the behaviour of the trap is analogous to that of a laser approaching the lasing threshold. These theoretical predictions are supported by the experimental results in Figs. (\ref{fig:subt}).

Time correlations (Supplementary Section) measure the rate at which the motion loses coherence due to thermal fluctuations . In general, $\langle x(t+\tau) x(t) \rangle$ and $\langle y(t+\tau) x(t) \rangle$ are damped oscillations. As $P\rightarrow P_0$, see Table (\ref{tab:quants}), the amplitude of $\langle x(t+\tau) x(t) \rangle$ increases as $1/(P_0-P)$, and decays increasingly slowly with time. Meanwhile, $\langle x(t+\tau) y(t) \rangle$ has the same amplitude, but is phase shifted by $\pi/2$, indicating a growing tendency towards rotation of the center of mass about the beam axis. 
Experimental results strongly support this idea, see Figs. (\ref{fig:subt}c,d). Once more, they clearly demonstrate an increase of motional coherence for circularly polarized beams for powers approaching the threshold power. 

For zero time ($\tau=0$), the instantaneous variances of the particle coordinates, e.g. $\langle x^2 \rangle$, quantify the departure from thermodynamic equilibrium. In a conservative, harmonic trap (i.e. linearly polarized beams) equipartition is satisfied and the elastic and thermal energies can be equated e.g. $\frac{1}{2}P k \langle x^2 \rangle = \frac{1}{2} k_BT$, so that the product, $P\langle x^2 \rangle$ is constant. 
In contrast, the circularly polarized trap deviates increasingly from equipartition as $P_0$ is approached, so that $\langle x^2 \rangle \propto 1/(P_0-P)$ increases rapidly, as confirmed experimentally in Fig. (\ref{fig:subt}e). 
Figure (\ref{fig:subt}f) illustrates this phenomenon visually:  the two-dimensional probability density distribution of the trapped particle in the transversal plane, scaled by the square root of the beam power, is invariant for a conservative trap but grows with increasing power in the presence of non-conservative forces.  

In combination, the PSD, time correlations and variances reveal the underlying physical behaviour. Optical forces vanish on the beam axis and are locally restoring so that points on the beam axis are mechanical equilibria. Nevertheless, thermal fluctuations repeatedly push the particle off axis, exposing it to the small, azimuthal spin forces. These increase the tendency for the particle to orbit about the beam axis, as expressed by the time correlations,  $\langle x(t + \tau) y(t) \rangle$ and $\langle x(t + \tau) x(t) \rangle$. Centripetal forces grow and act against the gradient forces, effectively cancelling them as $P \rightarrow P_0$. Under these conditions, the stochastic motion becomes increasingly coherent in terms of rotation: kinetic energy is increasingly concentrated at a single, resonant frequency as indicated in the PSD (Fig. (\ref{fig:subt}b)). Conversely, the radial motion is increasingly incoherent, and the particle is free to explore the trap (i.e. $\langle x^2 \rangle \rightarrow \infty$).

As the power is increased above $P_0$, centripetal forces tend to exceed gradient forces, and the particle is pushed outward. If the force field were truly linear, this process would continue indefinitely. In reality, the trap is finite. Stable orbits with well defined radii, $r_o$, and frequency, $\Omega$, can form as the particle encounters non-linear regions of the trap, taking us into the \textbf{high power regime II} (purple curves, Fig. (\ref{fig:beam+exp}d)). 
\section{Regime II: Stable Orbits}
Simple equilibrium conditions for the orbits can be obtained by balancing azimuthal spin forces with viscous drag, and radial centripetal forces with optical gradient forces:
 
\begin{subequations} \label{eq:oeqm}
\begin{align}
\frac{P}{\xi^2}&=-\frac{r_o}{m}\frac{{f_r} (r_o)}{f_{\phi}^2 (r_o)} \label{eq:oeqm_pwr},\\
\frac{\Omega}{\xi}&=\frac{{f_r}(r_o)}{m {f_{\phi}}(r_o)} \label{eq:oeqm_frq}.
\end{align}
\end{subequations}
Where $P$ is the optical power required to sustain an orbit with radius $r_0$ for a given viscous drag, $\xi$. Taking the linear approximation for small orbits (e.g. $f_i(r) \approx k_i r$, with $i=r,\phi$) in Eqs. (\ref{eq:oeqm}a,b), reproduces the threshold conditions between regimes I and II, Eqs. (\ref{eq:bal1},\ref{eq:bal2}) : the optical power required to zero the imaginary parts of the characteristic frequencies, i.e. $\Im(\omega_c(P))=0$, is precisely the power required for centripetal forces to balance gradient forces in the linear approximation. Because the ratio of the azimuthal and radial forces is approximately constant the orbit frequency, $\Omega$, is also approximately constant and independent of orbit radius and optical power.\\
In order for an orbit to be stable and coherent, further conditions must be met. Any change in the orbit radius or frequency should be counteracted by the trap. For instance, a perturbation in the radius will cause a change in the gradient, azimuthal and centripetal forces which should act to restore the orbit radius. Furthermore, these responses need to be strong enough to constrain thermal fluctuations. The linear stability of the orbits, and their resistance to thermal fluctuations are analysed in Supplementary Information. In summary, we find that for a beam with waist radius $W_0$, there are no asymptotically stable orbits with $r_0  \gtrsim r_{max} \equiv \sqrt{2/3} W_0$, which corresponds to a maximum optical power given by Eq. (\ref{eq:oeqm_pwr}) . Furthermore, thermal fluctuations tend to dominate orbits with small radii and a higher optical power ($P_{orb}>P_0$) is required for formation of thermally stable, coherent orbits (Supplementary Information).

\begin{figure}
	\includegraphics[width=0.9\textwidth]{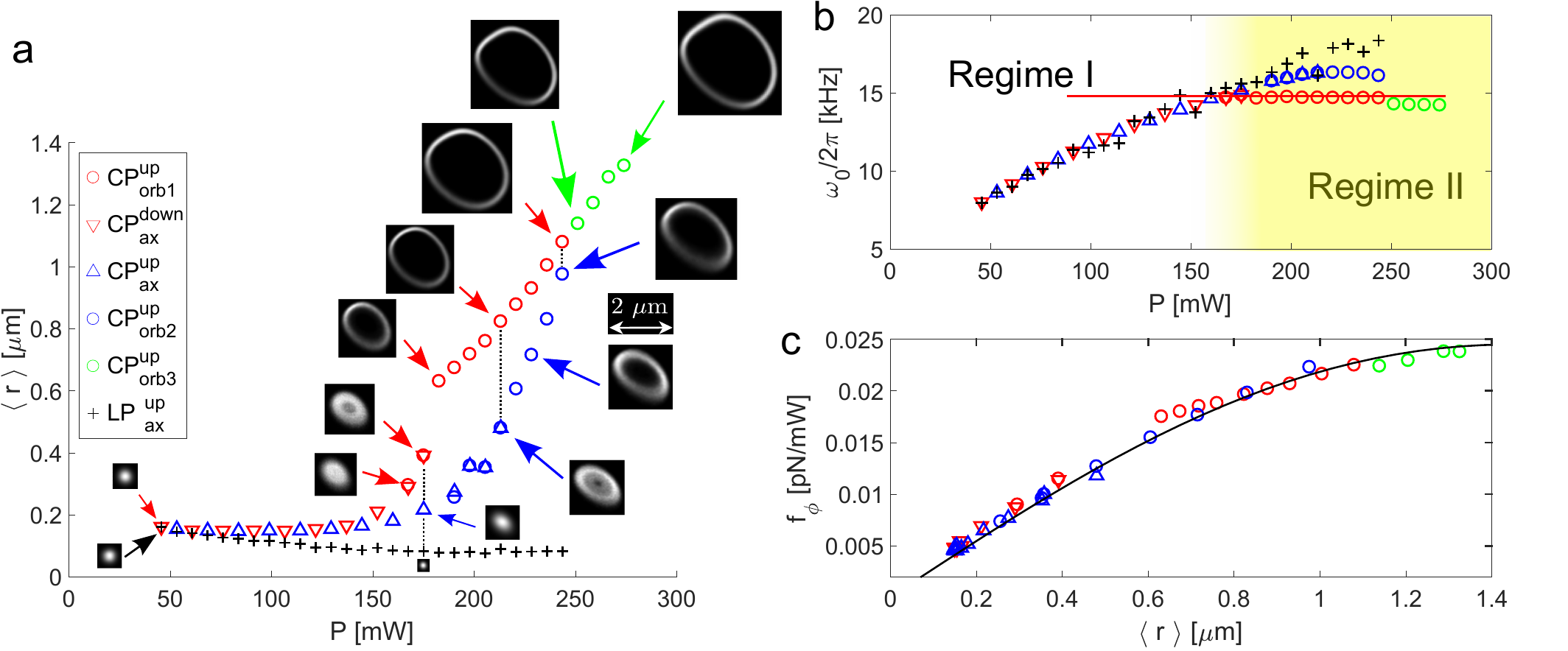}
	\caption{\label{fig:orb} Graphical comparison of the key physical quantities in regimes I and II. (a) Mean value of the orbit radius, $\langle r \rangle$, as a function of optical power. The insets show the probability density of the particle positions in the lateral $xy$ plane relative to the given scale bar. Data related to circularly polarized beams (CP) are denoted by {$\bigcirc$}, $\bigtriangleup$ and $\bigtriangledown$ for the orbiting regime II. $\bigtriangleup$ and $\bigtriangledown$ symbols correspond to increasing and decreasing optical power, respectively. For comparison data corresponding to parallel linear polarization (LP) are marked by black $+$ signs. The sequence in which the data was acquired is indicated by colors and markers {\color{red} {$\bigcirc$}}, {\color{red} {$\bigtriangledown$}}, {\color{blue} {$\bigtriangleup$}}, {\color{blue} {$\bigcirc$}}, {\color{black} {$+$}}, and {\color{green} {$\bigcirc$}}.  
	Overlapping points {\color{red} {$\bigcirc$}} with {\color{red} {$\bigtriangledown$}} and {\color{blue} {$\bigtriangleup$}} with {\color{blue} {$\bigcirc$}} illustrate how different calculation methods, used for regime I and II, overlap in the region where the orbiting is not fully developed.
	(b) Resonant trap frequency $\omega_0/2 \pi$ or orbiting frequency $\Omega/2 \pi$ for regime I or II, respectively. The theoretical value from Eq. (\ref{eq:oeqm_frq}), is indicated by the horizontal red line. Measured values of $\xi/m$, and calculated values of $f_r/f_{\phi}$ have been used.
	(d) Spin force $f_{\phi}$ for a particle of radius $770$ nm and density 2200 kg/m$^3$. The presented values were determined as the mean value for $x$ and $y$ directions from fits to the PSD (Supplementary Information ) for regime I and from the orbiting equation (Eq. \ref{eq:oeqm_frq}) for regime II (see details in Supplementary Information). The grey curve shows the theoretical force, obtained from generalized Lorentz Mie theory (Supplementary Information). A single scaling parameter has been used (Supplementary Information ).
 }
\end{figure}

This overall behaviour is confirmed experimentally. Figure (\ref{fig:beam+exp}c) illustrates the qualitative difference in the transverse motion of the particle between regimes I and II. This is quantified in Fig. (\ref{fig:orb}a), which shows the mean values of the radius of the orbit as a function of optical power. Insets show that the orbit is slightly elliptical and its size suffers hysteresis with respect to the direction of power change. In particular, orbiting appears to survive to lower powers when the power is being decreased. These effects are related to the difference between $P_0$, the power required to destabilize the trap in the linear regime, and $P_{orb} > P_0$, the power required to form a stable,  coherent orbit. At the lowest powers the mean radii of the particle trajectories are indeed comparable for CP and LP beams. However, they begin to diverge with increasing power; the LP radii decrease and CP increase, in accordance with theory. We note that the non-orbiting trajectories for the linearly polarized beams (black crosses in Fig. (\ref{fig:orb}a)) were captured between the orbiting trajectories shown by the blue and green circles, demonstrating the robustness of the orbiting behaviour. At higher optical powers than those represented in the graph, centripetal forces exceed gradient forces and the particle is radially ejected from circularly polarized traps.

Mean values for $x$ and $y$ trap frequencies ($\omega_0 /2\pi$) in LP traps are shown in Fig. (\ref{fig:orb}b) and, together with the CP results, they follow the predicted behaviour. In regime I, CP and LP traps follow the same power dependence, however in regime II the orbiting frequency remains constant, the exact value depending slightly on the branch of the hysteresis curve. This reflects the observation that the ratio of the radial and azimuthal forces is approximately constant, yielding a constant orbit frequency in Eq. (\ref{eq:oeqm_frq}).

Finally, Fig. (\ref{fig:orb}c) shows the measured azimuthal force as a function of radius. 

This data is evaluated by fitting to experimentally measured trajectories as described in the Methods section, below, and in Supplementary Information. In particular, the curve was evaluated by fitting experimental PSDs to theoretical profiles (Supplementary Information) for regime I, and by applying the orbiting equation, Eq. (\ref{eq:oeqm_frq}), in regime II. The validity of the theoretical description is further demonstrated by the close overlap of the data points in the transitional region between axial motion and fully developed orbiting. \\\\

\section{Discussion}
Our experiments vividly demonstrate the mechanical effects of optical spin momentum. Spin forces bias Brownian motion, induce orbits and ultimately throw a probe particle from an optical trap. These results are completely generic: for a given particle and circularly polarized beam, at a given air pressure, there is always an escape power above which the trap will eject the particle. For fixed optical power, there is also an equivalent pressure beneath which the same thing will happen. \\
In addition, this work provides an archetypal example of linearly non-conservative optical vacuum traps. All of the phenomena observed here arise from intrinsic, non-symmetric coupling between motional degrees of freedom. This is expressed in the asymmetry of the stiffness matrix, Eq. (\ref{eq:stiff}). For circularly polarized light, this coupling is caused by spin. More generally, asymmetric coupling can be induced by any reduction in symmetry, either of the particle shape, or the trapping field \cite{Simpson2010First-order}. In viscous media, this results in biased Brownian motion \cite{Irrera2016Photonic}. As we have seen, the consequences in air or weak vacuum are more extreme, since the trapped particle can accumulate momentum, resulting in thermally excited coherent motion. Analogous effects may be expected for non-spherical particles or defective beams.\\
Finally, these experiments signpost a new research direction. Optical force fields are generally non-conservative, and their structure is derived from internal momentum flows. It is clear that measuring the trajectories of probe particles in vacuum can help us develop our understanding of optical momentum flows in structured light fields as well as the dynamics and thermodynamics associated with motion in potentially complex force fields. Some of these possibilities have previously been described by Berry and Shukla \cite{Berry2013Physical,Berry2016Curl}. Nevertheless, major unanswered questions persist, especially concerning the details of momentum coupling to matter of varying shape and composition, and to the interplay between thermal fluctuations and some of the dynamical behaviour predicted elsewhere.

\section{Methods}
\subsection{Experimental set-up}

\begin{figure}
\includegraphics[width=0.9\textwidth]{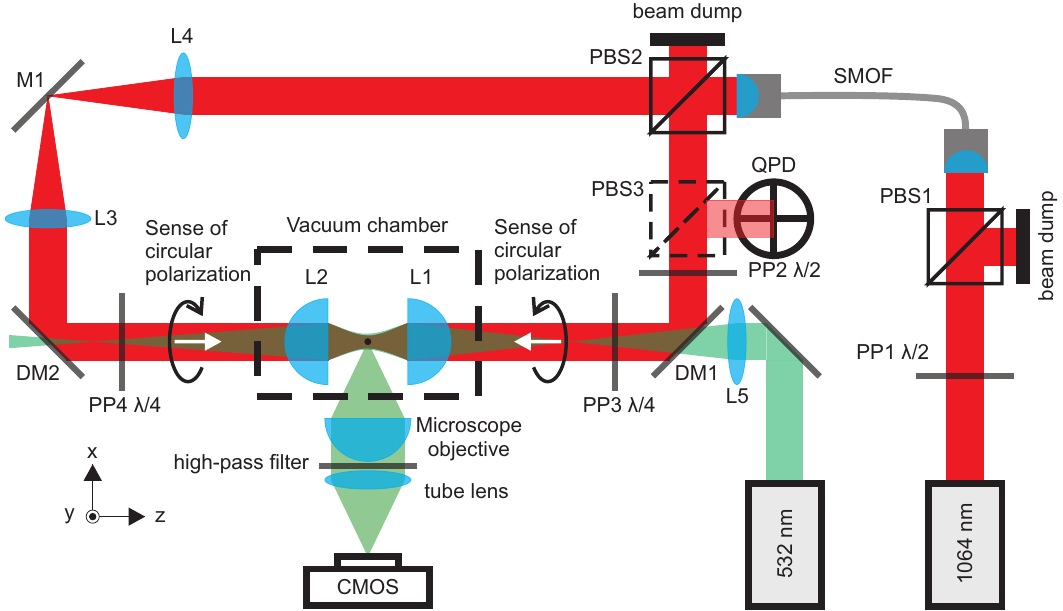}
\caption{\label{fig:exp} Schematic showing the optical set-up. A detailed description is provided in the Methods section.}
\end{figure}

A schematic of the experimental set-up is provided in Figure (\ref{fig:exp}). The low noise laser beam from a Prometheus laser (Coherent) working at a wavelength of $1064$nm passes, first, through a polarizing beamsplitter PBS1 and then is coupled to a single mode optical fiber (SMOF) for spatial filtering. The trapping power is controlled by rotation of the half-wave plate, PP1, in front of PBS1. The output of the SMOF is divided into two beams of equal power and orthogonal linear polarization (LP) by polarizing beamsplitter PBS2. Lenses L3 and L4 constitute a $1/1$ beam expander. Before entering the vacuum chamber, the polarization of one of the beams could be changed from linear to right-hand or left-hand circular (CP) by a quarter-wave plate PP4. The linear polarization of the second beam can be first rotated through an arbitrary angle by half-wave plate, PP2, and subsequently changed from linear to right-hand or left-hand circular polarization by the quarter-wave plate, PP3. In the vacuum chamber, the counter-propagating beams were focused by aspheric lenses L1, L2 (Thorlabs C240TME-C) of focal length $8$mm to create a trap with beam waist of $W_0 \approx 2.7 \mu m$. Non-porous SiO$_\mathrm{2}$ microspheres (Bangs Laboratories, diameter 1.54\,$\mu$m, size CV $10 - 15 \%$) with negligible absorption were sprayed with a nebulizer (Breuer IH50) into the volume of the optical trap at atmospheric pressure. Prior to spraying, the spheres were suspended in isopropyl alcohol with low particle concentration and the suspension was sonicated in an ultrasonic bath for $15$ minutes to prevent clumping. All subsequent experiments were performed at $4$ mbar. The pressure in the trap was measured with $30$\% accuracy with a Pirani and cold cathode gauge (Pfeiffer PKR 251). For CP measurements the two counter-propagating beams were circularly polarized with opposite handedness, so that the transverse electric fields of the beams rotate together (Fig. (\ref{fig:beam+exp}a)). Using the quarter wave plates we made comparative measurements with parallel or perpendicularly linearly polarized beams. The same particle was used for all the presented experiments which were performed with optical trapping power varying in the range $45$ - $281$ mW. 

\subsection{Particle Tracking} \label{subs:track}
Motion of the trapped particle was tracked using a quadrant photodetector (InGaAs QPD Hamamatsu G6849). We employed a detection scheme similar to back focal plane interferometry, where the scattered trapping light interferes with the unscattered trapping beam and creates a rotationally non-symmetrical intensity distribution in the plane of QPD \cite{PralleMRT99}. In our case, the trapping beam, propagating from left to right, was used as for the particle tracking at the QPD. While trapping in CP light, after passing through the vacuum chamber the measuring beam is reflected upwards out of the plane of the setup by the polarizing beamsplitter PBS3. On the other hand, while trapping in LP light, the measuring beam would have been transmitted by PBS3 so PP3 was rotated by 2$^\circ $ from its ideal orientation in order to get a signal on QPD. This allowed us to monitor the motion of the trapped sphere in axial ({\it z}) and orthogonal horizontal ({\it x}) and vertical ({\it y}) directions. The QPD signal was processed by home-made electronics and acquired by NI USB 6351 card at 250 kHz sampling rate. The particle’s motion was also tracked from outside the vacuum chamber by a fast CMOS camera (Vision Research Phantom V611) which was triggered to begin recording at the same time as the QPD. The microscope was composed of a microscope objective (MOTIC, Plan Apo, $100\times/0.55$, WD 13), tube lens (Thorlabs ACA-254-500A) and a fast CMOS camera (Vision Research Phantom V611). For the purpose of observation the particle was illuminated by a broad, weak laser beam of wavelength $532$nm, as the camera is more sensitive at this spectral range. The framerate was $200000$ fps for measurements in LP beams, when the particle remains close to the beam axis, and $187290$ fps for measurements of orbiting. The position of the particle in each frame was obtained with sub-pixel precision by fitting a two-dimensional Gaussian profile to the image \cite{Cheezum2001}. The microscope was calibrated using a calibration grid so the camera recording eventually provided calibrated {\it x} and {\it z} coordinates of the particle trajectory (nm). Finally, the calibration constant relating the QPD signal to displacement in meters was obtained for each power separately by comparing widths of histograms of position obtained by camera (nm) and by QPD (V). This comparison was possible because the camera and QPD recordings were triggered to start simultaneously. Since the CMOS camera detects $x$ and $z$ axis, we used the same calibration constant for $x$ and $y$ axes of QPD.

\subsection{Data processing}
Theoretical profiles were fitted to experimentally measured PSDs for LP beams, allowing us to determine the ratio of the drag to the mass of the particle, $\xi/m$. This was used as a fixed parameter for the system in analysis of particle motion in CP beams. Resonant trap frequencies $\omega_0$ or orbiting frequencies $\Omega$ were determined from fits to theoretical power spectra and the spin force contribution was obtained from fits to theoretical power spectra in regime I or using Eq. (\ref{eq:oeqm_frq}) for regime II. Further details are provided in Supplementary Information.

\bibliography{orbits}
\bibliographystyle{naturemag} 
\end{document}